\documentclass{iaus}
\usepackage{graphicx}

\title[Habitable Zones in  Multiple Planetary
Systems] 
{Habitable Zones for Earth-mass Planets in Multiple Planetary
Systems}

\author[Ji et al.]   
{Jianghui JI$^{1}$, Lin LIU$^2$,  Hiroshi  KINOSHITA$^3$ \and
Guangyu LI$^1$}

\affiliation{$^1$Purple  Mountain  Observatory, Chinese  Academy
of  Sciences,  Nanjing  210008, China \break email: jijh@pmo.ac.cn\\[\affilskip]
$^2$Department of Astronomy,  Nanjing University, Nanjing  210093,
China \break email: xhliao@nju.edu.cn\\[\affilskip]
$^3$National Astronomical Observatory,  Mitaka, Tokyo 181-8588,
Japan}

\pubyear{2007}
\volume{249}  
\pagerange{1--5}

\jname{Exoplanets: Detection, Formation and Dynamics}
\editors{Sylvio Ferraz-Mello, Yi-Sui Sun \& Ji-lin Zhou, eds.}
\begin{document}

\maketitle

\begin{abstract}
We perform numerical simulations to study the Habitable zones (HZs)
and dynamical structure for Earth-mass planets in  multiple
planetary systems. For example, in the HD 69830 system, we
extensively explore the planetary configuration of three
Neptune-mass companions with one massive terrestrial planet residing
in 0.07 AU $\leq a \leq$ 1.20 AU, to examine the asteroid structure
in this system. We underline that there are stable zones of at least
$10^5$ yr for low-mass terrestrial planets locating between 0.3 and
0.5 AU, and 0.8 and 1.2 AU with final eccentricities of  $e < 0.20$.
Moreover, we also find that the accumulation or depletion of the
asteroid belt are also shaped by orbital resonances of the outer
planets, for example, the asteroidal gaps at 2:1 and  3:2 mean
motion resonances (MMRs) with Planet C, and 5:2 and 1:2 MMRs with
Planet D. In a dynamical sense, the proper candidate regions for the
existence of the potential terrestrial planets or HZs are 0.35 AU $<
a < $ 0.50 AU, and 0.80 AU $< a < $ 1.00 AU for relatively low
eccentricities, which makes sense to have the possible asteroidal
structure in this system.

\end{abstract}

\firstsection 

\section{Introduction}
To date, over 260 extrasolar planets have been discovered around the
nearby stars within 200 pc (Butler et al. 2006; The Extrasolar
Planets Encyclopaedia\footnote{As of Nov. 8, 2007, see
http://exoplanet.eu/catalog.php and http://exoplanets.org/}) mostly
by the measurements of Doppler surveys and transiting techniques.
The increasing numbers of known extrasolar planets are largely
attributed to increasing precision in measurement techniques.
Observational improvements will likely lead to more substantial
discoveries, including: (1)diverse multi-planetary systems, of which
more than 20 multiple systems with orbital resonance or secular
interactions are already known;(2)low-mass companions around
main-sequence stars (so-called super-Earths), e.g., 55 Cancri
(McArthur et al. 2004), GJ 876 (Rivera et al. 2005), HD 160691
(Santos et al. 2004; Gozdziewski et al. 2007); (3)a true Solar
System analog, with several terrestrial planets, asteroidal
structure and a dynamical environment consistent with terrestrial
planets in the Habitable Zone (HZ) (Kasting et al. 1993) that could
permit the development of life, e.g., Gl 581 (von Bloh et al. 2007);
(4)a comprehensive census of a diversity of planetary systems, which
will provide abundant clues for theorists to more accurately model
planetary formation processes (Ida \& Lin 2004; Boss 2006).

Lovis et al. (2006) (hereafter Paper I) reported the discovery of an
interesting system of three Neptune-mass planets orbiting about HD
69830 through high precision measurements with the HARPS
spectrograph at La Silla, Chile. The nearby star HD 69830 is of
spectral type K0V with an estimated mass of $0.86 \pm  0.03
M_{\odot}$ and a total luminosity of  $0.60 \pm 0.03 L_{\odot}$
(Paper I), about 12.6 pc away from the Sun. In addition, Beichman et
al. (2005) announced the detection of a large infrared excess owing
to hot grains of crystalline silicates orbiting the star HD 69830
and inferred that there could be a massive asteroid within 1 AU.
Subsequently, Alibert et al. (2006) and Paper I performed lots of
calculations to simulate the system and revealed that the innermost
planet may possess a rocky core surrounded by a tiny gaseous
envelope. This planet probably formed inside the ice line in the
beginning, whereas the two outer companions formed outside the ice
line from a rocky embryo and then accreted the water and gas onto
the envelope in the subsequent formation process. Hence, it is
important for one to understand the dynamical structure in the final
assemblage of the planetary system (Asghari et al. 2004; Ji et al.
2005), and to investigate suitable HZs for life-bearing terrestrial
planets (Jones et al. 2005; Raymond et al. 2006; von Bloh et al.
2007; Gaidos et al. 2007) advancing the space missions (such as
CoRot, Kepler and TPF) aiming at detecting them, thus one of our
goals is to focus on the issues of the potential Earth-mass planets
in the system.

\section{Dynamical Structure and Habitable Zones in HD 69830 system}
Modern observations by  \textit{Spitzer} and \textit{HST} indicate
that circumstellar debris disks (e.g., AU Mic and $\beta$ Pic) are
quite common in the early planetary formation. Beichman et al.
(2006) used \textit{Spitzer} to show that $13 \pm 3\%$ of mature
main sequence stars exhibit Kuiper Belt analogs. They further point
out that the existence of debris disks is extremely important for
the resulting detection of individual planets, and related to the
formation and evolution of planetary systems. As mentioned
previously, Beichman et al. (2005) also provide clear evidence of
the presence of the disk in HD 69830. Subsequently Paper I's
best-fit orbital solutions were for three Neptune-mass planets with
well-separated nearly-circular orbits, which may imply that the
HD69830 system is similar to our Solar System in that it is
dynamically consistent with the possible presence of terrestrial
planets and asteroidal and Kuiper belt structures. Hence, it
deserves to make a detailed investigation from a numerical
perspective.

\begin{figure}
\includegraphics[height=1.8in, width=2.4in]{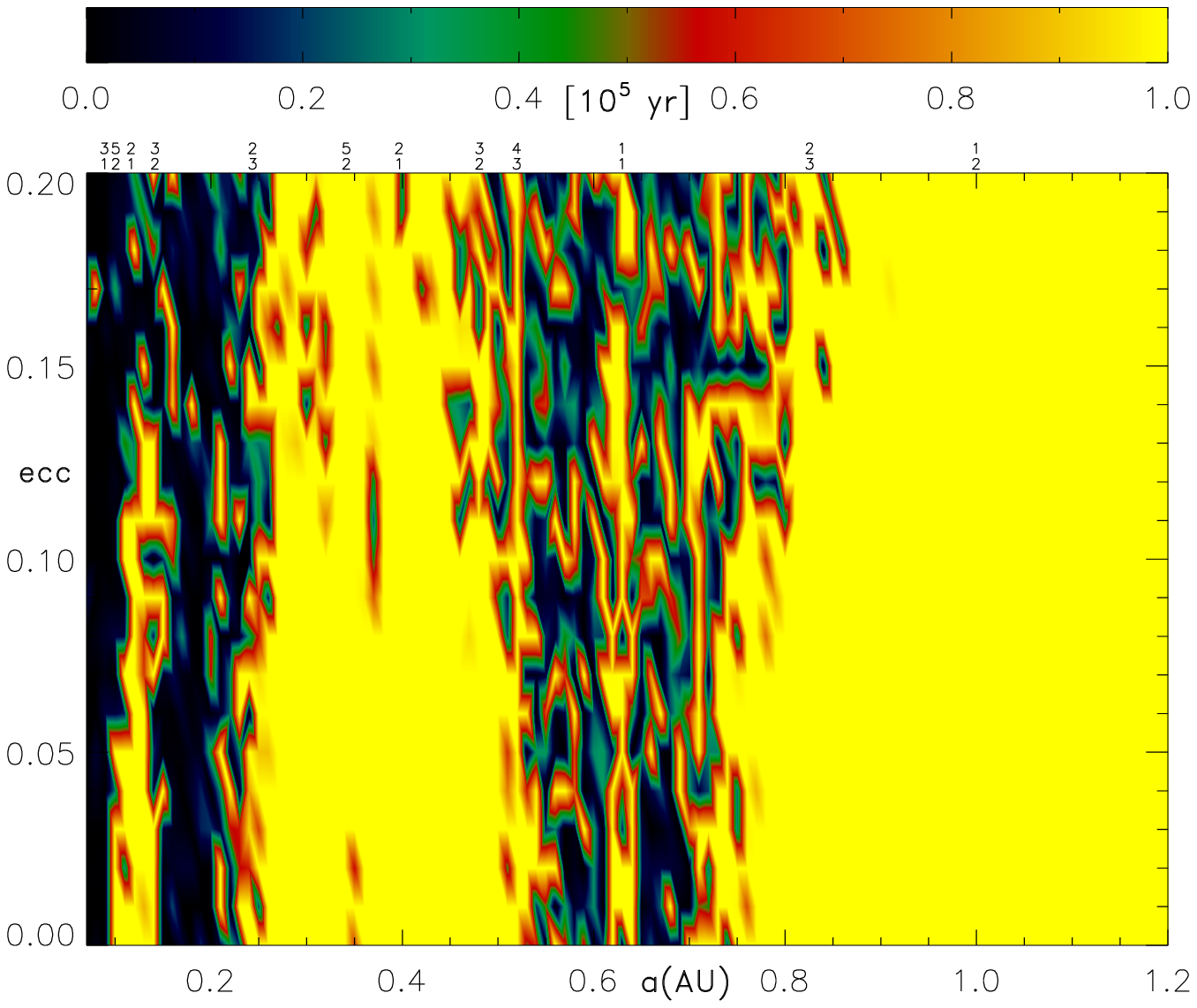}
\includegraphics[height=1.8in, width=2.4in]{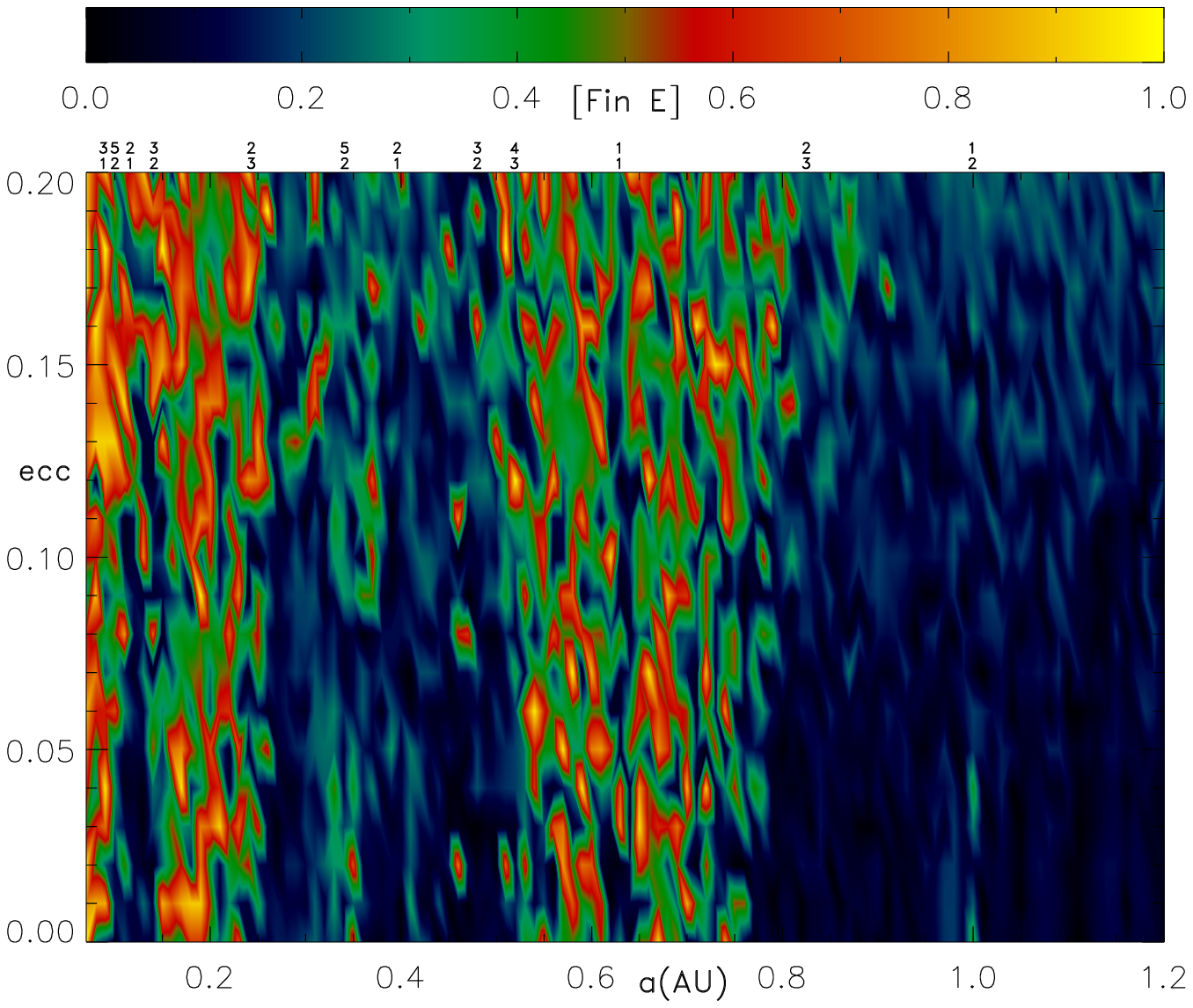}
  \caption{\textit{Left panel}:
  Contour of  the surviving time for Earth-like
  planets for the  integration  of  $10^5$ yr.
  \textit{Right panel}: Status of their final eccentricities.
  Horizontal and vertical axes are the initial  a and e.
  Stable zones for the low-mass planets in the region between 0.3 and
  0.5 AU, and 0.8 and 1.2 AU with final  low eccentricities.
} \label{fig1}
\end{figure}

To investigate the dynamical structure and potential HZs in this
system, we performed additional simulations with HD69830's three
Neptune-mass companions in coplanar orbits, and one massive
Earth-like planet. In the runs, the mass of the assumed terrestrial
planet ranges from 0.01 $M_{\oplus}$ to 1 $M_{\oplus}$. The initial
orbital parameters are as follows: the numerical investigations were
carried out in $[a, e]$ parameter space by direct integrations, and
for a uniform grid of 0.01 AU in semi-major axis (0.07 AU $\leq a
\leq $ 1.20 AU) and 0.01 in eccentricity ($0.0 \leq  e \leq 0.20 $),
the inclinations are $0^{0} < I < 5^{0}$, and the angles of the
nodal longitude, the argument of periastron, and the mean anomaly
are randomly distributed between $0^{0}$ and $360^{0}$ for each
orbit, then each terrestrial mass body was numerically integrated
with three Neptune-mass planets in the HD 69830 system. In total,
about 2400 simulations were exhaustively run for typical integration
time spans from $10^5$ to $10^6$ yr (about $10^6 - 10^7$ times the
orbital period of the innermost planet) (see also Ji et al. 2007 for
details).

Figure 1 shows the contours of the survival time for Earth-like
planets (\textit{left panel}) and the status of their final
eccentricities (\textit{right panel}) for the integration over
$10^5$ yr, where horizontal and vertical axes are the initial $a$
and $e$. The \textit{left panel} displays that there are stable
zones for a terrestrial planet in the regime between 0.3 and 0.5 AU,
and 0.8 and 1.2 AU with final eccentricities of $e < 0.20$.
Obviously, unstable zones exist near the orbits of the three
Neptune-mass planets where the planetary embryos have short
dynamical survival time, and their eccentricities can quickly be
pumped up to a high value $\sim$ 0.9 (\textit{right panel}). In
these regions the evolution is insensitive to the initial masses.
The terrestrial bodies  are related to many of the mean motion
resonances of the Neptunian planets and the overlapping resonance
mechanism (Murray \& Dermott 1999) can reveal their chaotic
behaviors of being ejected from the system in short dynamical
lifetime. Furthermore, most of terrestrial orbits are within
$3R_{hill}$ sphere of the Neptune-mass planets, and others are
involved in the secular resonance with two inner companions.

Analogous to our Solar system, if we consider the middle planet (HD
69830 c) as the counterpart as Jupiter, we will have the regions of
mean motion resonances: 2:1 (0.117 AU), 3:2 (0.142 AU), 3:1 (0.089
AU) and 5:2 (0.101 AU), 2:3 (0.244 AU). In Fig. 1, we notice there
indeed exist the apparent asteroidal gaps about or within the above
MMRs (e.g., 3:1 and 5:2 MMRs), while in the region between 0.10 AU
and 0.14 AU for $e < 0.10$, there are stable islands where the
planetary embryos can last at least $10^5$ yr. In addition, for
Planet D, most of the terrestrial planets in 0.50 AU $< a <$ 0.80 AU
are chaotic and their eccentricities are excited to moderate and
even high values, the characterized MMRs with respect to the
accumulation or depletion of the asteroid belt are 3:2 (0.481 AU),
2:1 (0.397 AU), 5:2 (0.342 AU), 4:3 (0.520 AU), 1:1 (0.630 AU), 2:3
(0.826 AU), 1:2 (1.000 AU), and our results enrich those of  Paper I
for massless bodies over two consecutive 1000-year intervals,
showing a broader stable region beyond 0.80 AU. Note that there
exist stable Trojan terrestrial bodies in a narrow stripe about
0.630 AU, involved in 1:1 MMR with Planet D, and they can survive at
least $10^6$ yr with  resulting small eccentricities in the extended
integrations.  The stable Trojan configurations may possibly appear
in the extrasolar planetary systems (see also Dvorak 2007; Psychoyos
\& Hadjidemetriou 2007 in this issue), e.g., Ji et al. (2005)
explored such Trojan planets orbiting about 47 Uma, and Gozdziewski
\& Konacki (2006) also argued that there may exist Trojan pair
configurations in the HD 128311 and HD 82943 systems. Ford  \& Gaudi
(2006) developed a novel method of detecting Trojan companions to
transiting close-in extrasolar planets and argue that the
terrestrial-mass Trojans may be detectable with present ground-based
observatories.  Terrestrial Trojan planets with low eccentricity
orbits close to 1 AU could potentially be habitable, and are worthy
of further investigation in the future.

\section{Summary and Discussion}
In this work, we investigated the planetary configuration of three
Neptune-mass companions similar to those surrounding HD 69830 and
added one massive terrestrial planet in the region of 0.07 AU $\leq
a \leq$ 1.20 AU to examine the dynamical stability of terrestrial
mass planets and to explore the asteroid structure in this system.
We show that there are stable zones of at least $10^5$ yr for the
low-mass terrestrial planets located between 0.3 and 0.5 AU, and 0.8
and 1.2 AU with final eccentricities of $e < 0.20$. Moreover, we
also find that the accumulation or depletion of the asteroid belt is
also shaped by orbital resonances of the outer planets, for example,
the asteroidal gaps of 2:1 and 3:2 MMRs with Planet C, and 5:2 and
1:2 resonances with Planet D. On the other hand, the stellar
luminosity of HD 69830 is lower than that of the Sun, thus the HZ
should shift inwards compared to our Solar System. In a dynamical
consideration, the proper candidate regions for the existence of the
potential terrestrial planets or HZs are 0.35 AU $< a < 0.50 $ AU,
and 0.80 AU $ < a < 1.00 $ AU for relatively low eccentricities.
Finally, we may summarize that the HD 69830 system can possess an
asteroidal architecture resembling the Solar System and both the
mean motion resonance (MMR) and secular resonances will work
together to influence the distribution of the small bodies in the
planetary system. In other simulations, we also show the potential
Habitable zones for Earth-mass planets in the 47 UMa planetary
system (see Ji et al. 2005), and the results imply that future
space-based observations, e.g., CoRot, Kelper and TPF will hopefully
produce a handful of samples belonging to the category of the
terrestrial bodies.

\begin{acknowledgments}
We thank the anonymous referee for informative comments and
suggestions that helped to improve the contents. This work is
financially supported by the National Natural Science Foundations of
China (Grants 10573040, 10673006, 10203005, 10233020) and the
Foundation of Minor Planets of Purple Mountain Observatory.
\end{acknowledgments}

\end{document}